\documentclass[english,10pt,rmp,twocolumn]{revtex4}
\usepackage{ae,aecompl}
\usepackage[latin9]{inputenc}
\usepackage{amsmath,amssymb,amsbsy}
\usepackage{graphicx}
\usepackage{fancyhdr}

\usepackage[letterpaper,margin=3cm,]{geometry}

\makeatletter
\usepackage{babel}
\makeatother

\begin{document}

\title{Embedding the Bilson-Thompson Model in a LQG-like framework}

\author{Deepak Vaid}

\date{\today}

\begin{abstract}
We argue that the Quadratic Spinor Lagrangian approach allows us to approach the problem of forming a geometrical condensate of spinorial tetrads in a natural manner. This, along with considerations involving the discrete symmetries of lattice triangulations, lead us to discover that the quasiparticles of such a condensate are tetrahedra with braids attached to its faces and that these braid attachments correspond to the preons in Bilson-Thompson's model of elementary particles. These "spatoms" can then be put together in a tiling to form more complex structures which encode both geometry and matter in a natural manner. We conclude with some speculations on the relation between this picture and the computational universe hypothesis. 
\end{abstract}

\maketitle

\tableofcontents

\section{Introduction}\label{sec:intro}

LQG and String Theory both remain a few steps away from giving a coherent description of quantum gravity which naturally incorporates the particles of the SM - i.e. the so-called goal of "Unification". However, we have obtained a very good notion of what the final picture should look like from the advances in the respective fields. In fact now we are faced with a convergence of two supposedly clashing approaches. Critics of String Theory point to its lack of a natural habitat for the SM and its many solutions constituting an embarrasment of riches that is yet to be tamed. It doesn't have one indisputable conclusion or equation, but a plethora of very compelling ideas which, it is safe to say, will emerge naturally in the final analysis. Likewise the main weakness of LQG, when viewed as a candidate theory for unification, its lack of a particle spectrum, does not diminish the validity of the physical implications of quantum geometry.

Given the abundance of theoretical evidence, it is clear that any notion of particles as topological structures should find a natural home in LQG and String Theory, and should have a realization as particles of a QFT obtained by coarse-graining over the microscopic degrees of freedom.

In 2006 Sundance Bilson-Thompson proposed proposed just such a model \cite{Sundance2005Topological} for the particles of the Standard Model (SM), or at least those in the first generation: the leptons consisting of the electron, electron-neutrino and the up and down quarks and the gauge bosons ($W^\pm$, $Z_0$, $\gamma$) could be given a unified representation in terms of the irreducible elements of the first non-trivial braid group ($B_3$).\footnote{To be precise, Bilson-Thompson used an enlargement of the braid group.  Physically this consists of replacing the 1D threads of the braid with 2D ribbons which can then contain twists (or orientation). Mathematically this is the product group $\tilde{B}_3 = B_3 \times Z_2$ - i.e. the product of the simplest abelian and the simplest non-trivial braid group.} He then showed that the irreducible elements of $\tilde{B}_3$ can be put into one-to-one correspondence with (at least) the first generation of the SM particles in a very natural manner. Despite the elegance of the construction - for instance all particles have left and right-handed representatives, except for the neutrino which comes in only one handedness - some significant physical questions remained unanswered in \cite{Sundance2005Topological}. In the following we elaborate on these missing pieces.

Now, at least at a purely visual level, the braid picture seems to be in concordance with the structures that are natural in both LQG and ST - Spin-networks whose 1D edges can braid around each other\footnote{Indeed, Yidun Wan has shown \cite{Wan2007Braid} that this process allows us to implement a variant of Bilson-Thompson's picture, on graphs embedded in a three-dimensional topological space.} on the one hand, and 1D strings and higher dimensional brane-like structures on the other. Unfortunately, this visual similarity begins and ends at the purely speculative level and can only guide us to the final answer. It has yet to be shown how to correctly embed ribbon-like structures in LQG. Smolin has stated \cite{Smolin2002Quantum} that in LQG with a positive $\Lambda$, for reasons involving the regularization of n-point functions of the gauge field, we are required to use framed ribbons instead of 1D curves as the edges of our spin-networks.



The plan of the rest of the paper is as follows. In section \ref{sec:topological_considerations} we review the status of our understanding of elementary particles in the frameworks of LQG and String Theory and present physical motivation for the mathematical constructions. In section \ref{sec:qsl} We introduce the Quadratic Spinor Lagrangian (QSL) formulation of the gravity action which allows us to treat the tetrads as matter fields. We argue that the term corresponding to a non-zero cosmological constant in the Einstein action has a natural interpretation as a four-fermion interaction when we identify tetrads with spinor fields. In section \ref{sec:braids} we argue that braids emerge naturally as maps between faces of tetrahedra when the discrete symmetries of a triangulation are taken into account. In section \ref{fig:particle_lattice} we illustrate an application of our model to the construction of more complex structures containing both geometry and matter, by tiling these \emph{spatoms}\footnote{We thank Sundance Bilson-Thompson for this terminology}. In section \ref{sec:comparison} we compare our construction to the work by Wan and collaborators and finally conclude with some observations and ideas for future work in section \ref{sec:outlook}.

metric degrees of freedom are encoded in the twisting and braiding of ribbons connecting different faces of a triangulation $\Delta \mathcal{M}$.


\section{Topological Considerations}\label{sec:topological_considerations}

LQG is born out of a 3+1 decomposition of the Einstein action (when formulated in connection and tetrad variables) yielding the Hamiltonian (H) of General Relativity (without matter). This Hamiltonian is found to be the sum of constraints (as would be expected to be the case for a diffeomorphism invariant theory) called the Gauss ($C_g$, Diffeomorphism $C_d$ and Hamiltonian $C_h$ constraints respectively. These constraints are functions of the connection $A^i_a$ and triad $e^a_i$ living on a 3-manifold $\Sigma$ which is a slice of the "complete" four dimensional spacetime on which the action was originally defined. In the general case $\Sigma$ can have internal boundaries (topologically two spheres $S^2$) and/or external ones (corresponding to a cosmological horizon). This will be our arena: A 3-dim manifold with arbitrary number of internal boundaries which are topologically $S^2$ and an outer boundary which would correspond to a cosmological (deSitter) horizon.

The highly successful application of the methods of connections and fiber bundles on manifolds to the description of the electroweak and strong forces naturally led physicists to explore whether such a description could encompass gravity as well. For if we could cast gravity as a gauge theory, then all the formidable methods and insights weaned from connections theories - such as Yang-Mills and Pati-Salam-Glashow-Weinberg - could then be used to attack the problem of quantizing gravity. In this direction a seminal breakthrough was made in 1988 by Ashtekar \cite{Ashtekar1986New, Ashtekar1987New} who managed to cast the equations of GR into a form where the basic variables were no longer the metric or the Christoffel connection (both highly unweildy mathematically), but a connection $A^i_\mu$ living on a background manifold accompanied by \emph{tetrad/vielbien} fields $e^\nu_i$ - which encode the usual metric\footnote{The relationship being given by: $g_{\mu\nu} = |e|\eta^{ij} e^i_\mu e^j_\nu$, where $\eta$ is the flat minkowski tensor} but whose behaviour under gauge transformations is much simpler than that of the metric\footnote{Under an arbitrary co-ordinate transformation (which is precisely the gauge degree of freedom of pure GR) $g_{\mu\nu} \rightarrow g'_{\alpha\beta}=\frac{\partial x^\mu}{\partial x'^\alpha}\frac{\partial x^\nu}{\partial x'^\beta} g_{\mu\nu}$ where $x'$ are the new co-ordinates expressed as functions of the old ones. Tetrads transform as: $e^{\mu} _{i} \rightarrow U e^{\mu}_{i}$, where $U$ is an element of the relevant gauge group}. The Ashtekar variables allow us to write an action for GR which is quadratic in the generalized momenta i.e. the tetrads, radically simplifying the complexity of the problem which previously contained second derivatives of the metric rendering it unamenable to standard analytic methods involving perturbation theory. This discovery, along with later work by Smolin, Rovelli, Pullin and others gave birth to the new field of Loop Quantum Gravity which is today the only viable alternative to String Theory as far as unification is concerned \footnote{There are other promising avenues but broadly speaking they fall into two categories: bottom-up models such as causal dynamical triangulations (CDT) and spin-foams, and top-down models such as string theory and modifications of GR}

Previous work in Lattice QCD had already shown us how to quantize a theory with a non-abelian gauge group, with the connection living on 1D lines of the edges of a graph and the fermions living at the vertices of a graph. This insight was exported to LQG, where along with certain characteristics peculiar to the gravitational case, it evolved into the present picture of space-time as being constructed of a (spin)foam like structure from which a classical geometry should emerge via a suitable coarse-graining. This approach however has run into an impasse in recent years. The results of quantum geometry and the counting of Black Hole microstates while impressive in their own right do not allow us to shove under the carpet - so to speak - the problem of matter. In LQG as it stands today, there is no natural way to incorporate the particles of the Standard Model. Yidun Wan's construction \cite{Wan2007Braid} along with further improvements \cite{BilsonThompson2006Quantum, BilsonThompson2008Particle} holds the promise of overcoming this obstacle. In this paper we show how this can be accomplished. It is here that topology comes into play. The internal boundaries we use replace the usual vertices of spin-networks\footnote{This idea, while discovered independently by the author, was mentioned in \cite{Smolin1995Linking} and is also used in Wan's construction \cite{Wan2007Braid} and in a recent proposoal by Rovelli and Krasnov \cite{Rovelli2009Black}}. The edges of graphs are replaced by framed ribbons which, as we will show below, fit neatly onto the surfaces of the internal boundaries. It also turns out that the braiding which is essential to describe the SM particles, emerges naturally on consideration of the action of discrete symmetries of the internal boundaries.

Over the past two decades an increasing amount of circumstantial evidence has accumulated for what is known as the Holographic Principle. First advocated by t'Hooft and Susskind, and later recast by Maldacena as the AdS/CFT conjecture, the claim is that what we perceive as physics in 3+1 dimensions is actually happening on 2+1 dimensional surfaces. Thus the world as we perceive is not 4-dimensional, but rather just as a three dimensional image can be reconstructed by shining light through a the two-dimensional hologram, the 4-dimensional Lorentzian physics we encounter can be reconstructed from interactions occurring on 2-dimensional surfaces. Any theory of Quantum Gravity which claims to include the particles of the SM as elementary excitations thus must somehow incorporate this paradigm. As we shall see the fundamental degrees of freedom in our construction live on 2-dim surfaces which are precisely the inner boundaries of $\Sigma$.





Aside from these reasons, topology has played an important role in many areas of physics. In high-energy physics monopoles and instantons are topologically non-trivial solutions of the Yang-Mills action. In the integer and fractional quantum hall effects, the states for which the hall conductivity vanishes correspond to topological solutions of the Chern-Simons action, labelled by the level $k$ for the IQHE and by fractional values for the FQHE. In nematic crystals and in crystal lattices, defects are characterized by topological invariants. In 2+1 quantum gravity, the local degrees of freedom correspond to topological defects (punctures) in the two dimensional spatial manifold.

Having given what we hope are satisfactory reasons for the topological nature of our arena we proceed with our construction. We start with showing how the gravitational action on manifolds with boundaries decomposes into a bulk and boundary term, the latter being identical to Chern-Simons theory.

\section{Quadratic Spinor Lagrangian for GR and Four-fermion interaction}\label{sec:qsl}

In \cite{Alexander2006Gravity,Alexander2007Fine} it was argued that cosmological inflation could be explained as occuring due to the formation of a BCS condensate of Majorana fermions. The unsatisfactory part of these works was that matter had to be introduced by hand. In this section we show that there exists a way to cast the Einstein action in a form in which the comological constant term can be naturally interpreted as corresponding to a four-fermion interactions between spinorial tetrads. Thus there is no need to insert matter fields by hand.

Consider a 4-manifold $\mathcal{M}$ of signature $+1$ with holes.
The boundaries of these holes are topologically $S^{2}\times \mathbf{R}$. 
On $\mathcal{M}$, there lives a gravitational connection $A_{\mu}^{I}$
a one-form which takes values in the $\mathfrak{sl}(2,\mathbf{C})$
lie-algebra and a tetrad $e^{\mu}_{I}$. $\mu,\nu,\dots$ stand for spatial indices and $I,J,\dots$ are lie-algebra indices. In terms of these variables the action for GR can be written as:

\begin{equation}\label{eqn:GRAction}
	S_{GR} = \int_{\mathcal{M}}d^{4}x\; \tilde{e}_{I} \wedge \tilde{e}_{J} \wedge R^{IJ}
\end{equation}

where $\tilde{e}^{\mu}_{I} = det(e) e^{\mu}_{I}$ are density weight $1/2$ objects. The $\wedge$ product is over the space-time indices which we have suppressed in the action to avoid clutter. Now consider the following \cite{Tung1995Spinor, Jacobson1988Fermions}. The curvature is given by:

\begin{equation}\label{eqn:Curvature}
	R^{IJ} = d A^{IJ} + g A^{IK} \wedge A_{K}^{J} 
\end{equation}

where $g$ is the gauge coupling. Then we have:

\begin{eqnarray}\label{eqn:QuadraticTetradIdentity}
	\mathcal{D}\tilde{e}_{\nu}^{I} & = & d\tilde{e}^{I} + g A^{I}_{J} \wedge \tilde{e}^{J} \\ \nonumber
\mathcal{D}^{2} \tilde{e}^{I} & = & d^{2} \tilde{e}^{I} + g d ( A^{I}_{J} \wedge \tilde{e}^{J} ) \\ \nonumber
										       &   & + \, g A^{I}_{J} \wedge (d\tilde{e}^{J} + g A^{J}_{K} \wedge \tilde{e}^{K}) \\ \nonumber
									& = &  g (d A^{I}_{J}  + g A^{I}_{K} \wedge A^{KJ}) \wedge \tilde{e}^{J} \\ \nonumber
									& = & g R^{IJ} \wedge \tilde{e}_{J}
\end{eqnarray}

Now substituting the last line of the above into (\ref{eqn:GRAction}) we get:

\begin{equation}\label{eqn:QuadraticTetradAction}
	S_{GR} = -\int_{\mathcal{M}} d^{4}x\;  g\,\tilde{e}_{I} \wedge \mathcal{D}^{2} \tilde{e}^{I} 
\end{equation}

Doing an integration by parts in the above equation yields:

\begin{equation}\label{eqn:QuadraticTetradAction2}
	S_{GR} = \int_{\mathcal{M}} d^{4}x \; g\, \mathcal{D}\tilde{e}^{I} \wedge \mathcal{D}\tilde{e}_{I} - \int_{\partial \mathcal{M}} d^{3}x \; g\, \tilde{e}^{I} \wedge \mathcal{D} \tilde{e}_{I}
\end{equation}

In \ref{eqn:QuadraticTetradAction2} the boundary term looks similar to the action for Chern-Simons theory. This similarity can be made more apparent by the inclusion of the cosmological term in the action:

\begin{eqnarray}\label{eqn:QuadraticActionWithLambda}
	S_{GR \Lambda} & = & S_{GR} + S_{\Lambda} \\ \nonumber
						& = &  \int_{\mathcal{M}} d^{4}x \; g\, \mathcal{D}\tilde{e}^{I} \wedge \mathcal{D}\tilde{e}_{I} - \int_{\partial \mathcal{M}} d^{3}x \; g\, \tilde{e}^{I} \wedge \mathcal{D} \tilde{e}_{I} \\ \nonumber
						& & 	- \int d^{4}x\; \frac{\Lambda}{det(e)} \, \tilde{e}^{I} \wedge \tilde{e}^{J} \wedge \tilde{e}^{K} \wedge \tilde{e}^{L} \epsilon_{IJKL}
\end{eqnarray}

Now the last term must be evaluated in the bulk and also on the boundary. Consequently $S_{GR\Lambda}$ takes the form:

\begin{equation}
	S_{GR\Lambda} = S_{\mathcal{M}} + S_{\partial \mathcal{M}}
\end{equation}

where:

\begin{eqnarray}
	S_{\mathcal{M}} & = & \int_{\mathcal{M}} d^{4}x \; [ g\, \mathcal{D}\tilde{e}^{I} \wedge \mathcal{D}\tilde{e}_{I} \\ \nonumber
							   &  & - \frac{\Lambda}{det(e)} \, \tilde{e}^{I} \wedge \tilde{e}^{J} \wedge \tilde{e}^{K} \wedge \tilde{e}^{L} \epsilon_{IJKL}] \\
	S_{\partial \mathcal{M}} & = & - \int_{\partial \mathcal{M}} d^{3}x \; [ g\, \tilde{e}^{i} \wedge {}^{3}\mathcal{D} \tilde{e}_{i} \\ \nonumber 
							   &  & + \Lambda \, \tilde{e}^{i} \wedge \tilde{e}^{j} \wedge \tilde{e}^{k} \epsilon_{ijk} ]
\end{eqnarray}

where the last term is corresponds to the Chern-Simons theory of spinors $e_{a}^{i} \sigma^{a}$ living on the boundary, where $\sigma_{a}$ are the Pauli matrices\footnote{As an aside we also want to note that the bulk action resembles that of the non-linear sigma model (NLSM). This analogy points to a possible way of understanding the non-trivial fixed points of GR via the methods which have been applied to the NLSM \cite{Hamber2009Quantum-Gravity}}.

Now the physical picture of matter we have is that of a sea of defects living on many internal boundaries which are ``floating`` in the bulk manifold. In this case we have multiple disconnected boundaries and the boundary action is more appropriately written as:

\begin{equation}\label{eqn:BoundaryAction}
	S_{\partial \mathcal{M}} = \sum_{m \in \mathcal{S}} S^{m}_{\partial \mathcal{M}}
\end{equation}

i.e. the sum of the Chern-Simons action on all the boundaries in $\mathcal{M}$. 

The offshoot of all of this is to provide some mathematical justification for our somewhat heuristic construction in what follows. Casting the gravitational action in this way illuminates the duality between matter and geometry that is already present in the classical theory. Ordinarily, gravity is understood to be the interaction between geometry, as represented by metric variables, and matter as represented by Standard Model fields. The expression (\ref{eqn:QuadraticTetradAction2}), however, shows us an alternate perspective in which matter is already present in the form of the tetrad field. The cosmological constant term in (\ref{eqn:QuadraticActionWithLambda}) can then be seen to correspond to a four-fermion interaction. Now, as is well understood, the presence of such a term implies that the ordinary vacuum with $<\tilde{e}^{I}\tilde{e}^{J}> = 0$ is unstable and starting with a free fermionic gas with such an interaction, a symmetry breaking transition to a BCS ground state occurs in which $<\tilde{e}^{I}\tilde{e}^{J}> \ne 0$.

\begin{figure}[htp]
\centering
\includegraphics[scale=0.20]{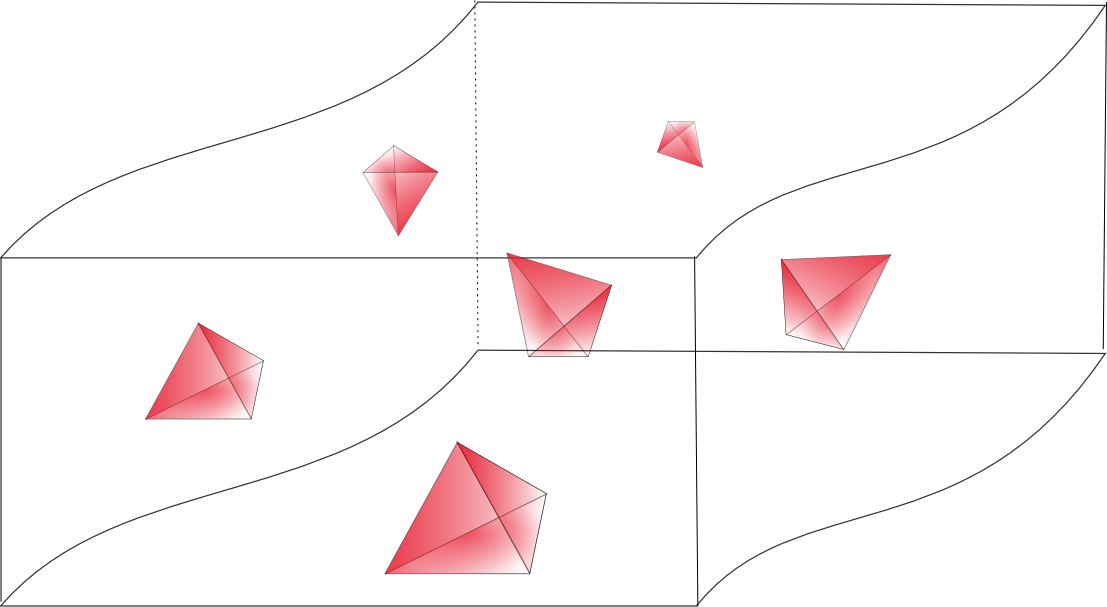}
\caption{Tetrahedra floating in space with topology $\mathbb{R}^{3}$}\label{fig:floating_tetrahedra}
\end{figure}

Thus we are led to the following picture of pre-geometry: an abstract 3 dimensional topological space with a fermionic sea, which undergoes spontaneous symmetry breaking leading to the formation of a condensate, where the natural interpretation of the resulting quasi-particles is that of quanta of geometry. The resulting situation is depicted in Fig. \ref{fig:floating_tetrahedra}. Now, a priori, we have no reason to proceed further. We are left with a gas-like state of geometry\footnote{This leads us to speculate that one will have different phases of geometry characterized by the amount of connectivity of the tetrahedra.}. Such a state is unlikely to yield an approximate continuum geometry because we have no natural measure on this space of tetrahedra. What we require is a graph-like picture as in the spin-network construction. For this purpose the tetrahedra must somehow be connected to each other. In the process of trying to discover the right way to glue these tetrahedra together we are naturally led to the emergence of the Bilson-Thompson model. That is the topic of the next section.

\section{Braids on inner boundaries}\label{sec:braids}

The crucial missing element in Bilson-Thompson's original construction is an explanation of just \emph{where} these braids live. In LQG we have a manifold without a fixed background metric. Are these braids just to be visualized as floating in this pre-geometric manifold much like proteins in the primeval sludge? In the absence of a metric, how are we to have a notion of locality? It turns out that the additional structure needed in order to answer these questions is the presence of internal boundaries within $\mathcal{M}$.

It has been rigorously shown in \cite{Ashtekar2000Quantum} that in canonical LQG, the state space of the punctures on the surface of an isolated horizon (i.e. inner boundary) corresponds to that of Chern-Simons theory. In this regard see also \cite{Engle2009Black, Rovelli2009Black}. In a separate paper we argue why the punctures living on the isolated horizon can be considered to be fermionic (or more generally anyonic) degrees of freedom which undergo condensation to form quanta of geometry. The interpretation of topological punctures, characterized by a deficit angle $\theta$, as particles is consistent with the results of \cite{Ashtekar2000Quantum} and is also in line with various proposals for realizing matter as topological structures in QG \cite{Alexander2009Superconducting, Alexander2003Geometrization}. While this assumption is not yet based on a rigorous mathematical footing, we feel that the consequences of this approach present a natural avenue for investigation and that is what we attempt in the following.

\subsection{Triangulations of a 2 sphere}

Most viable schemes for constructing a theory of quantum geometry exploit the partition function approach. This involves replacing the bulk continuous manifold by a discrete one constructed from simplices whose faces, edges and vertices are labelled by various spins and operators whose values determine the quantum state of geometry for each simplex. For details see \cite{Perez2005Introduction}. When applied to a manifold with internal boundaries, one has to discretize the boundary manifold also. For a 3+1 manifold, the bulk contains 4-simplices. The 2+1 dimensional boundaries are then covered with 3-simplices - i.e. tetrahedra. Now in general the boundary surface will have space-like, time-like and null patches. These distinctions require the presence of a metric. For a topological theory it is only the \emph{topology} that matters which in this case we take to be $S^{2} \otimes R$.
 
\begin{figure}[htp]
\centering
\includegraphics[scale=0.3]{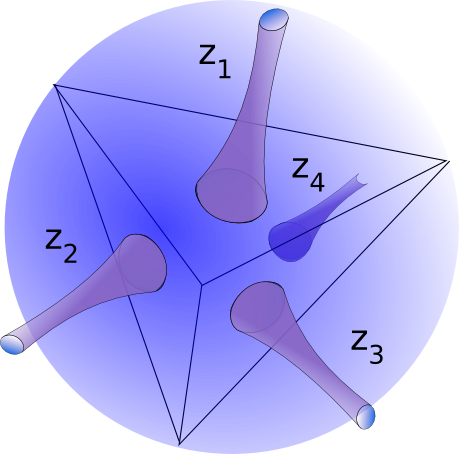}
\caption{A triangulated 2-sphere surface punctured by flux tubes labelled by stereographic co-ordinates $z_{i}$ }\label{fig:PuncturedSphere}
\end{figure}

So now we focus on a hole in a three dimensional manifold, whose surface ($S^2$) is punctured by lines carrying gravitational flux Fig. \ref{fig:PuncturedSphere}. The smallest non-trivial triangulation of a 2-sphere requires four triangles - giving us a tetrahedral approximation to the surface. The reason for a minimum of four punctures is simple. The surface of the sphere can be stereographically mapped to the complex plane $\mathbb{C}$. Ordinarily two field configurations on $S^2$ would be physically equivalent as long as they are related by an $SL(2,\mathbb{C})$ transformation\footnote{This being the induced action on $S^2$ of Lorentz transformations of external observers [cite Penrose, Rindler]}. Now it is a well-known result of projective geometry that given two sets of three (or fewer) points on $S^2$ we can always find a conformal map which takes one set into the other. Thus any set of three punctures can be mapped onto a circle thus depriving us of the topology of $S^2$. However a set of four points can be mapped into another set of four points, iff both sets have the same cross-ratio.\footnote{where the cross ratio for four points ${z_1,z_2,z_3,z_4}$ on $\mathbb{C}$ is $\lambda = \frac{z_1 - z_3}{z_2 - z_3} \frac{z_1 - z_4}{z_2 - z_4}$}. As a result, the specification of four punctures on $S^2$ and their co-ordinates ${z_1,z_2,z_3,z_4}$, under a given stereographic projection, breaks the continuous symmetry of the complex plane from the M$\ddot{\textrm{o}}$bius Group (PSL(2,$\mathbb{C}$)) to the Modular Group (PSL(2,$\mathbb{Z}$)), elements of which preserve the cross-ratio.\footnote{Physically, this means that a state of the sphere with four punctures transforms for external observers according to SL(2,$\mathbb{Z}$) instead of SL(2,$\mathbb{C}$), i.e. Lorentz transformations are quantized}

We imagine each face our tetrahedron $\Delta$ is pierced by one flux line, endowing that face with an area. The enclosed volume represents the smallest irreducible atom of geometry.\footnote{The conclusion that the vertices in the standard spin-network picture should be replaced by tetrahedra has also emerged via a different line of reasoning in recent work by Freidel, Krasnov, Livine and others \cite{Freidel2009Holomorphic}} To see how this comes about we refer the reader to [cite: condensate paper]. This is our point of departure from the standard LQG framework.

\subsection{Mapping between Faces of Tetrahedra}

Our playground is an arbitrary 3-manifold populated with these elementary quanta of volume represented by tetrahedra. The face of each tetrahedron is punctured by a flux line whose other end terminates either on another tetrahedron, or dangles in the bulk as an unattached node representing a free fermion.

We can use symmetry arguments to elucidate the structure of these flux tubes. The diffeomorphism gauge freedom of the bulk translates into the conformal invariance of the gauge fields living on the 2-spheres. Once we establish the triangulation, however, the conformal symmetry is broken. It now applies on the individual faces but not on the simplex as a whole. A single continuous patch can no longer cover the simplex, which is a piecewise linear manifold - or in less technical terms, the gauge fields defined on a face are not differentiable at the edges and vertices.

\begin{center}
\begin{figure}[htp]
\includegraphics[scale=0.2]{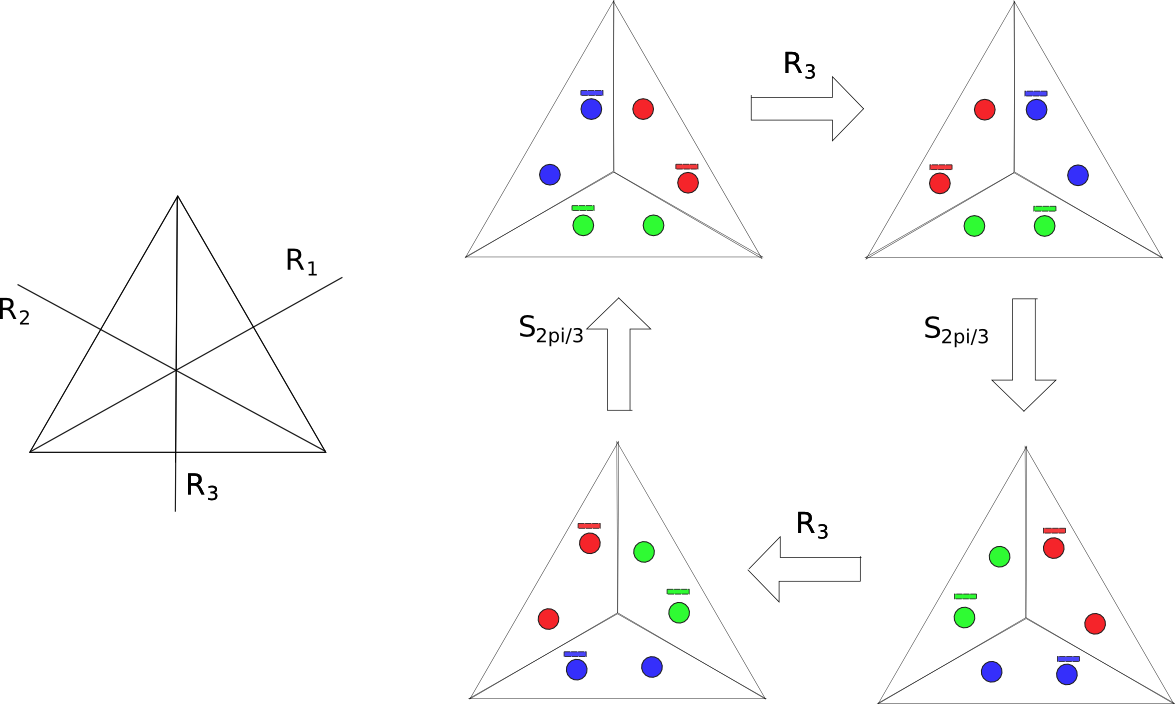}
\caption{Representation of the action of elements of the dihedral group ($D_2$) on a triangle. The triangle on the right hand side shows the axes of reflection. Thus $R_i$ stands for reflection across the $i^{\textrm{th}}$ axis of symmetry. The generators of rotations are $S_\theta$, with positive rotations being in the counterclockwise direction}\label{fig:DihedralGroup}
\end{figure}
\end{center}

The correct symmetry to work with in this situation is the discrete symmetries of the triangulation. For each triangle this corresponds to the dihedral group $D_{2}$ - consisting of the reflections, rotations and permutations . The generators of $D_{2}$ are a 3-fold rotation around the transverse axis, and three reflections as illustrated in Fig. \ref{fig:DihedralGroup}

\begin{figure}[htp]
\centering
\includegraphics[scale=0.3]{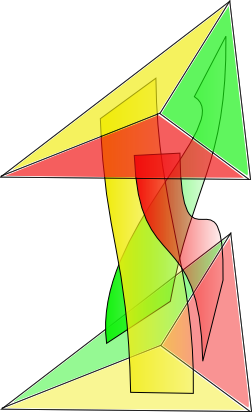}
\caption{Map between triangle represented by an element of the braid group }\label{fig:TriangleMap}
\end{figure}

Thus we see that the intersection of the flux tubes with the 2-sphere is no longer topologically a circle. The circle must be broken up into three segments in order to accommodate the discrete symmetries of the triangulation.

\subsection{Physical Interpretation of Torsion}

For a manifold $M$ equipped with a connection $\omega_{\mu\nu}^I{}_J$ and vielbien $e_{\mu}{}^J$, the torsion tensor is defined as the exterior covariant derivative of the vielbien with respect to the connection $\omega$, as in:

\begin{equation}
	T_{\mu\nu}{}^I = d_{[\mu}e_{\nu]}{}^I + \omega_{[\mu}^I{}_J e_{\nu]}^J \nonumber
\end{equation}

\noindent or suppressing the manifold indices:

\begin{equation}
	T^I = d\, e^I + \omega^I{}_J \wedge e^J = \mathbf{d}_{[\omega]} \, e^I
\end{equation}

Clearly, torsion measures the vorticity or circulation of the vielbien. This is most evident in 2 dimensions. There the vielbien becomes a dyad ${\mathbf{e}_x,\mathbf{e}_y}$ in terms of local co-ordinates. Now the torsion $T_{ab}$ is a 2-form written explicitly as:

\begin{equation}
	T_{ab} = \left( \begin{array}{cc} 0 & \partial_{[x}e_{y]} \\
							\partial_{[y}e_{x]} & 0 \\ 
			\end{array} \right)
				+ \omega \wedge e
\end{equation}

The 2D torsion tensor can be identified with the conductivity tensor of a 2D quantum hall system. In analogy with the quantum hall effect, we expect that like the conductivity tensor, the torsion will also be quantized. In terms of braids, the torsion measures the amount of \emph{twisting} a ribbon connecting two triangular faces as in \ref{fig:TriangleMap} undergoes. In the Bilson-Thompson model these ribbons are required to undergo a complete twist i.e. by $2\pi$ in the CCW or CW sense. The amount of the twist and its handedness then corresponds to the magnitude and sign of the charge carried by a ribbon.

\section{Tiling Spacetime}\label{sec:tiling}

\begin{figure}[htp]
\centering
\includegraphics[scale=0.1]{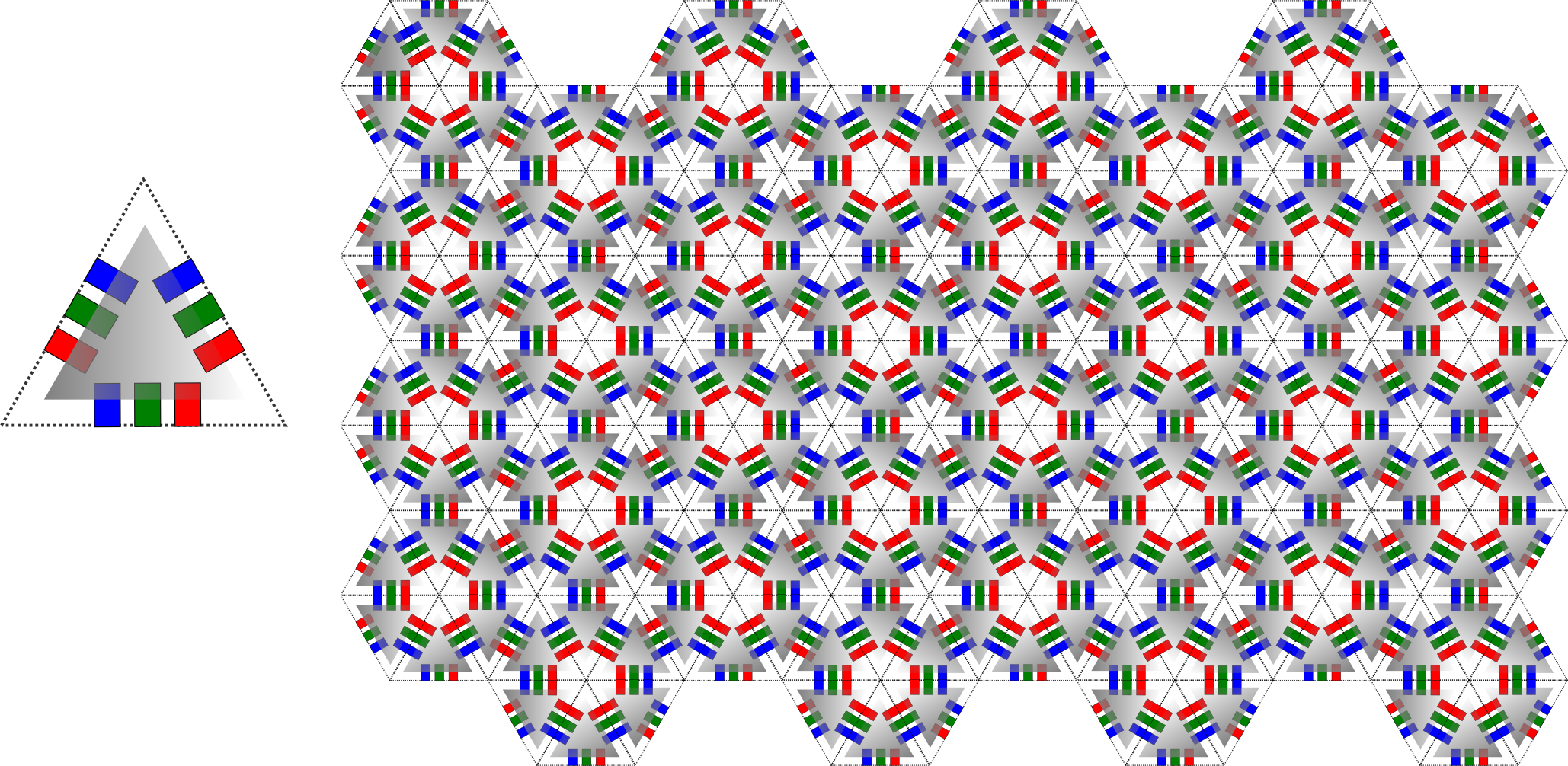}
\caption{A tiling generated via application of discrete symmetries to "spatoms" represented by the elementary tile on the left.}\label{fig:particle_lattice}
\end{figure}

The mathematical theory of tiling has long tackled a subject of great fascination to some people but of seemingly little useful application in the real world. Take for instance, Penrose tiles - an irreducible set of tiles which tile the two dimensional flat space aperiodically, i.e in a pattern which does not exhibit any ODLRO. Such sets of tiles would certainly make for beautiful bathrooms, but is there more to their beauty than meets the eye? In Fig. \ref{fig:particle_lattice} we present a hexagonal tiling generated by a single tile, demonstrating the possibilities inherent in our model. The tiling is generated by application of \textbf{P3M1}\footnote{Corresponding to a reflection of the tile, followed by a rotation by 120$^\circ$. This tiling was generated by using the "clone tool" of the image manipulation program \emph{Inkscape} freely available from http://www.inkscape.org/}, which is one element of the so-called wallpaper groups, to the single tile shown on the left of Fig. \ref{fig:particle_lattice}. 

The represents a single "spatom" with preons attached to three faces of a tetrahedron. The fourth face lies in the plane of the paper, with its preon attachment either coming out of or going into the paper. For simplicity, we consider only two-dimensional tilings by leaving the fourth face unattached, and we ignore details of the braiding between the ribbons, representing this aspect by the different colors of the ribbons.

\begin{figure}[htp]
\centering
\includegraphics[scale=0.1]{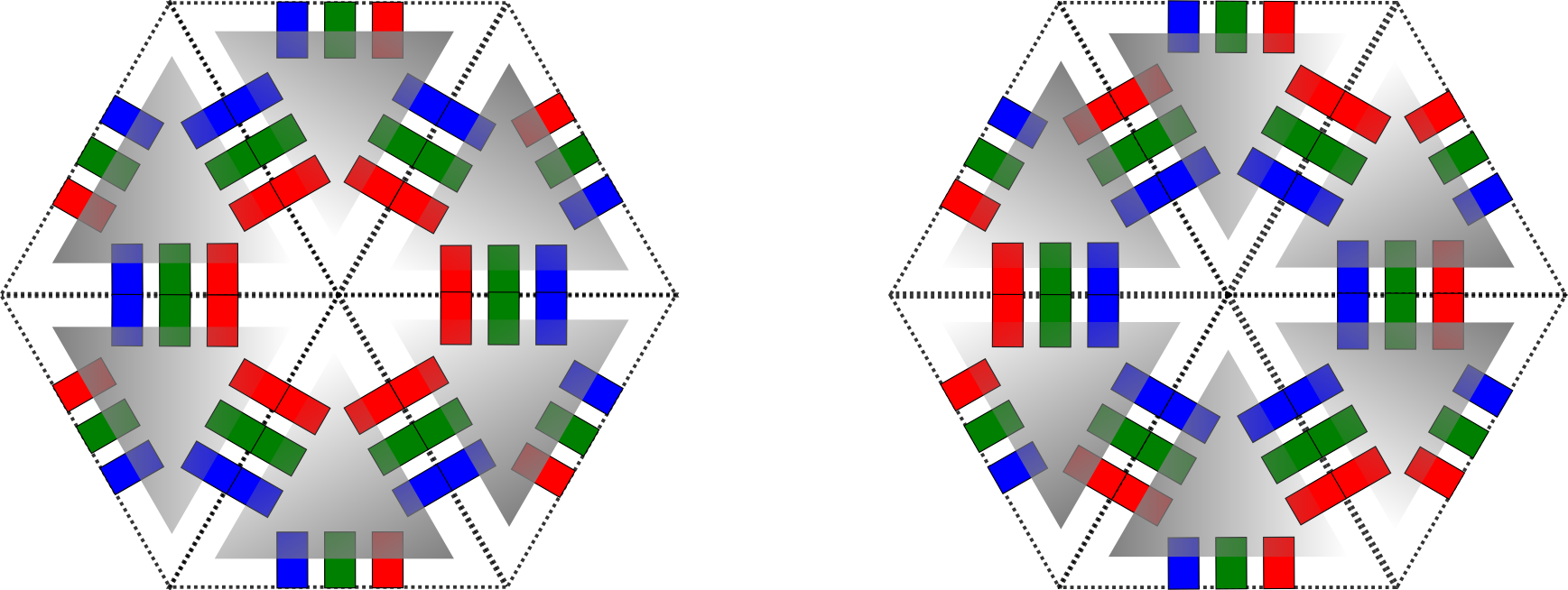}
\caption{Two examples of emergent excitations obtained from the elementary tile}\label{fig:emergent_excitation}
\end{figure}

The resulting two-dimensional lattice shows, on closer inspection, the excitations which can emerge from such a model. These emergent degrees of freedom, shown in Fig. \ref{fig:emergent_excitation} correspond to the closed loops formed by the red and blue ribbons in the tiling in Fig. \ref{fig:particle_lattice}.

%

\section{Comparison to the Wan Model}\label{sec:comparison}

As mentioned earlier, work along very similar lines has been done by Yidun Wan and collaborators including Lee Smolin, Fotini Markoupoulu, Jonathan Hackett and Sundance Bilson-Thompson. While there are some similarities between Wan's model (which we shall refer to as \textbf{I}) and ours (referred to as \textbf{II}), there are some subtle differences that distinguish the two.

The common motivation behind \textbf{I} and \textbf{II} is to find some way to embed Bilson-Thompson's preons into a framework of quantum geometry. Both models assume that ribbon-like objects are attached to the surfaces of tetrahedra, which would then play the role of the quanta of geometry. The crucial difference lies in the structure of these attachments and in how the braiding is implemented. In \textbf{I} a tube is attached to each face of a tetrahedron. The twists required for charges then emerge from applying the rotational symmetry of a triangle to which a given tube is attached. As for the braids, they are put in by hand, so to speak, in that there isn't any simple symmetry that generates the braiding between different strands. In \textbf{II} the braiding occurs between three ribbons which are attached onto a \emph{single} face of a tetrahedra. This picture arises due to the need to preserve the symmetries of the gauge fields living on the faces under dihedral rotations and reflections of the same. Consequently, given three unbraided and untwisted ribbons attached to a face, under the action of the elements of the dihedral group on that face, one generates braiding and twisting between these ribbons. Thus in \textbf{I} four tubes meet at a vertex, and the preons correspond to a braiding between three of those tubes. In \textbf{II} on the other hand each face can have three ribbons piercing it, corresponding to four elementary preons meeting at a given vertex, as in the elementary tile shown in Fig. \ref{fig:particle_lattice} (the fourth face is in the plane of the paper).

\begin{figure}[htp]
\centering
\includegraphics[scale=0.30]{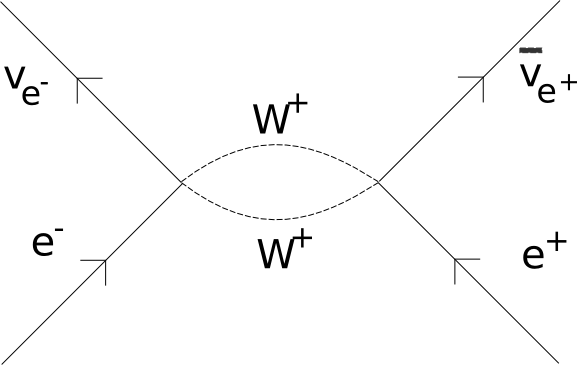}
\caption{Non-renormalizable, dimension 6 vertex for the process $e^{-} + e^{+} \rightarrow \nu_{e} + \bar \nu_{e} $}
\label{fig:JJWW_vertex}
\end{figure}

Apart from the technical differences, there is also a different physical motivation and implication of \textbf{II}. Our model emerged from considerations involving the formation of a geometrical condensate of spinorial tetrads as shown in section \ref{sec:qsl} . Therefore in \textbf{II} the tetrahedra stand for quanta of geometry. In \textbf{I} there is no clear identification of tetrahedra as geometrical quanta\footnote{Though, to be fair, \textbf{I} is built by replacing the vertices of graphs tetrahedra, the identification of which with geometrical quanta is all but unavoidable. Also our construction is no less heuristic.}. Whereas in\textbf{I} a preon can be associated only with one face of a given vertex; in our model, for each face of a tetrahedron there is one preon - correspondingly at each vertex four preons can meet. This picture seems somewhat more natural and elegant when applied to the problem of interactions between preons in order to regularize interactions vertices in QFT. For instance consider the following QFT vertex in Fig. \ref{fig:JJWW_vertex}

\begin{figure}[htp]
\begin{center}
\includegraphics[scale=0.4]{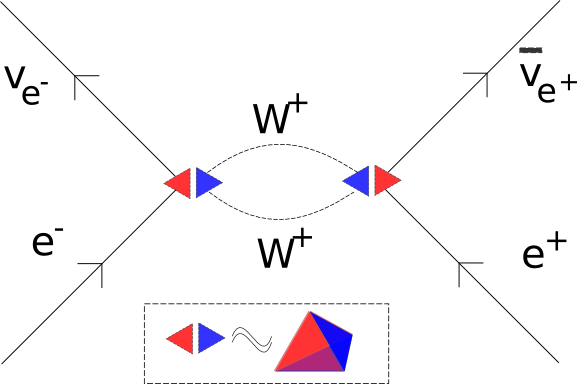}
\caption{Finite vertex for $e^{-} + e^{+} \rightarrow \nu_{e} + \bar \nu_{e} $ obtained by replacing interaction vertices by tetrahedra. For ease of illustration we represent the tetrahedra by two triangles as indicated in the box.}
\label{fig:FiniteVertex}
\end{center}
\end{figure}

This process has a divergent amplitude because it corresponds to dimension 6 operator. As mentioned in the introduction, there has been an expectation that an understanding of quantum geometry would allow us to regularize such pathological interactions. In our model 3-strand braids representing the fermions of the standard model can be joined onto faces of tetrahedra representing quantum geometry, suggesting a regularization of the vertex (\ref{fig:JJWW_vertex}) in the form of \ref{fig:FiniteVertex}.

Now instead of summing over momenta, one would sum over the space of shapes of tetrahedra compatible with the external braid attachments. This observation combined with the recent result that the holomorphic factorization of the quantum tetrahedron yields the CFT 4-vertex amplitude of string theory \cite{Freidel2009Holomorphic},  leads us to the possibility of a direct relationship between string theory and quantum geometry. This will be the subject of future work.

\section{Outlook}\label{sec:outlook}

Einstein's equation:

\begin{equation}
	G_{\mu\nu} = 8 \pi G\, T_{\mu\nu} - \Lambda g_{\mu\nu}
\end{equation}

and its connection to quantum physics has mystified physicists for generations. It is generally accepted that in a quantum description geometry and matter will be on an equal footing. There will be, in some sense, a duality between matter and geometry allowing us to map quantities in strong field regions to those in a dual weak-field spacetime. The inspiration from this idea came from the notion of \emph{composite fermions} \cite{Jain2007Composite, Jain1989Compositefermion, vonKlitzing1986Quantized} in condensed matter systems. There, the argument is that in a two dimensional electron gas (2DEG) in the presence of strong magnetic fields, $p$ electrons pair up with $q$ quanta of magnetic flux, resulting in quanta with fractional charge $p/q$. In this way a system of strongly-coupled particles is transformed to one where the new particles are weakly-interacting in a lower background effective field.

The identification of representations of the braid group with elementary particles also has strong implications for the \emph{computational universe} paradigm. Such considerations
are suggested by the recent advances in quantum computation particularly as implemented
in a two dimensional electron gas (2DEG). If a strong connection between
the physics of black holes and the QHE can be established, then it
would become less controversial to make arguments in support of the
computational universe paradigm. Indeed, if there exists a mapping
between processes in the QHE and BH physics then this would naturally
lead to the notion that the structure of space-time being essentially
topological in nature (i.e. constructed of braids, loops, strings
etc.) can be thought of as hardware, the software corresponding to
which would be topological rules governing physical processes such
as the scattering of elementary particles at the most fundamental
level.

It also happens to be the case that in a recent work Kauffmann and Lomonaco \cite{Kauffman2004Braiding} have shown that the elements of $B_{3}$ act as universal gates for quantum computation. The description of two fundamental aspects of nature, computation and particle physics, in terms of identical topological structures must be more than a merry coincidence. This connection remains to be explored further.



\begin{acknowledgements}
I would like to thank Sundance Bilson-Thompson and the Perimeter Institute for hosting me in September 2009. I also thank Sundance, Gregory Chaitin, Joao Ochoa, Lee Smolin and Stephon Alexander for helpful discussions and comments. This work is dedicated to my parents and my family whose love and support made it possible.
\end{acknowledgements}

\bibliography{bib_library}

\end{document}